\documentclass[fleqn,usenatbib]{rasti}

\usepackage[T1]{fontenc}
\usepackage{amsmath}
\usepackage{amssymb}
\usepackage{booktabs}
\usepackage{graphicx}
\usepackage{microtype}
\usepackage{placeins}
\usepackage{url}
\hypersetup{hypertexnames=false}
\IfFileExists{newtxtext.sty}{%
  \usepackage{newtxtext,newtxmath}%
}{%
  \usepackage{mathptmx}%
}
\usepackage{orcidlink}

\DeclareRobustCommand{\VAN}[3]{#2}
\let\VANthebibliography\thebibliography
\def\thebibliography{\DeclareRobustCommand{\VAN}[3]{##3}\VANthebibliography}
\newcommand{\Nninety}{N_{90}}
\newcommand{\prt}{\texttt{petitRADTRANS}}
\newcommand{\mcs}{Mission Candidate Sample}
\newcommand{\chisq}{\chi^{2}}
\title[Ariel Tier-2 requirements beyond 5H]{Beyond Five Scale Heights: Composition- and Cloud-dependent Ariel Tier-2 Requirements for sub-Neptunes}
\author[V. Krishnamurthy et al.]{
Vigneshwaran Krishnamurthy\,\orcidlink{0000-0003-2310-9415}$^{1}$\thanks{E-mail: vignesh.krishnamurthy@astr.tohoku.ac.jp},
Nicolas B. Cowan\,\orcidlink{0000-0001-6129-5699}$^{2,3}$,
Misato Fukagawa\,\orcidlink{0000-0003-1117-9213}$^{1}$ and
Teruyuki Hirano\,\orcidlink{0000-0003-3618-7535}$^{4,5}$
\\
$^{1}$Astronomical Institute, Graduate School of Science, Tohoku University, 6-3 Aoba, Aramaki-aza, Aoba-ku, Sendai 980-8578, Japan\\
$^{2}$Department of Earth \& Planetary Sciences, McGill University, 3450 rue University, Montr\'eal, QC H3A 0E8, Canada\\
$^{3}$Department of Physics, McGill University, 3600 rue University, Montr\'eal, QC, H3A 2T8, Canada\\
$^{4}$Astrobiology Center, 2-21-1 Osawa, Mitaka, Tokyo 181-8588, Japan\\
$^{5}$National Astronomical Observatory of Japan, 2-21-1 Osawa, Mitaka, Tokyo 181-8588, Japan\\
}
\date{Submitted for RASTI Ariel Special Issue; comments/suggestions are welcome}
\pubyear{\the\year{}}
\begin{document}
\label{firstpage}
\pagerange{\pageref{firstpage}--\pageref{lastpage}}
\maketitle
\begin{abstract}
The Ariel Mission Candidate Sample (MCS) Tier-2 requirements provide a common screening reference by assuming a clear H$_2$/He atmosphere and a five-scale-height transmission signal. We quantify how those requirements change for 197 known planets with $1.5\leq R_{\rm p}/R_\oplus<4.0$ when composition and clouds are specified explicitly. Equilibrium-chemistry transmission spectra at $1\times$, $10\times$, and $100\times$ solar metallicity are evaluated for clear atmospheres and idealized 1-mbar cloud decks, binned to the 47 AIRS Tier-2 channels, and combined with target-specific MCS-calibrated noise curves. For each atmosphere, we calculate the number of transits required for a featureless spectrum to be rejected at a $3\sigma$ confidence in at least 90\% of repeated observations. The clear $1\times$-solar atmosphere closely follows the MCS prediction. At $10\times$ solar metallicity, stronger molecular opacity generally improves detectability, whereas at $100\times$ solar metallicity the increased mean molecular weight dominates and roughly doubles the required number of transits. Cloudy atmospheres follow the same non-monotonic trend but at substantially greater cost: the $1\times$-solar, 1-mbar case requires nearly an order of magnitude more transits than the MCS prediction, while the enriched cloudy cases still require several times more. Overall, in most atmospheric scenarios tested, small planets require more transits to robustly detect spectral features than listed in the MCS. A literature crossmatch further identifies Ariel targets with detected metastable helium, for which Tier-2 molecular spectra could connect lower-atmosphere composition and clouds to an extended, escaping H$_2$/He upper atmosphere.

\end{abstract}
\begin{keywords}
Numerical methods -- planets and satellites: atmospheres -- space vehicles: instruments -- techniques: spectroscopic
\end{keywords}

\section{Introduction}\label{sec:introduction}
The Ariel mission will conduct a broad, homogeneous survey of exoplanet atmospheres to investigate planetary formation, evolution, and atmospheric diversity \citep{Tinetti2018,EdwardsTinetti2022}. Its tiered survey architecture trades sample size against spectral precision. Tier~2 provides moderate-resolution retrieval-quality spectra intended to reveal the principal molecular absorption bands for a large sample of planets, so the number of repeated observations assigned to each target affects both individual-planet science and population-level leverage \citep{Morales2022,Cowan2025}.

The Ariel \mcs{} (MCS) is regularly updated and provides estimates of the number of observations required to satisfy the requirements of each survey tier \citep{Edwards2019,EdwardsTinetti2022,MCSRepository}. These estimates use wavelength-dependent instrumental uncertainties calculated with ArielRad \citep{Mugnai2020}. For transmission observations, \citet{Edwards2019} adopted a mean atmospheric molecular weight of $\mu=2.3$, appropriate for a primordial H$_2$/He-dominated atmosphere. The current MCS documentation states explicitly that the required number of observations is calculated by requiring the Ariel medium-resolution InfraRed Spectrometer (AIRS) to reach $\mathrm{S/N}>7$ on the modulation expected from five atmospheric scale heights at the spectral resolution of the corresponding tier \citep{MCSRepository}. This uniform reference provides an efficient basis for comparing gas giants, but it may be optimistic for small planets with high-mean-molecular-weight secondary atmospheres or muted spectral features. The MCS documentation therefore cautions that such planets may require additional observations.

Recent work has emphasized that Ariel spectral information content depends on atmospheric state and on the adopted tier precision \citep{Radica2026}. Here we ask a complementary question: how much does the AIRS Tier-2 requirement for a small planet change when the atmosphere is enriched or cloudy rather than the clear H$_2$/He reference used for MCS screening?

\section{Sample and Methods}\label{sec:methods}

\begin{figure}
\centering
\includegraphics[width=\columnwidth]{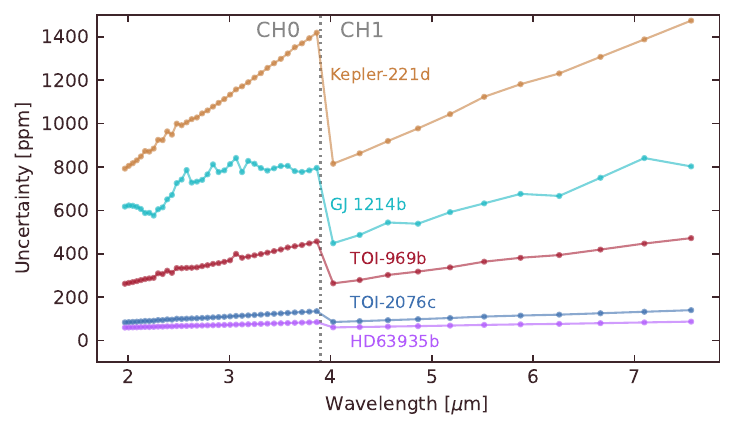}
\caption{Representative uncertainty spectra for Ariel/AIRS after a single transit. The vertical dotted line marks the AIRS-CH0/AIRS-CH1 boundary at $3.9\,\micron$. These curves retain wavelength-dependent target weighting but are not independent ArielRad simulations.}
\label{fig:noise_curves}
\end{figure}

We calculate equilibrium-chemistry transmission spectra for 197 known planets in the MCS with $1.5 \leq R_{\rm p}/R_\oplus < 4.0$ across six atmospheric scenarios, and evaluate the complete AIRS spectral structure. The scenarios combine three metallicities: $1\times$, $10\times$, and $100\times$ solar with either no clouds (``clear'') or an idealized opaque gray cloud deck at 1 mbar. For each planet and scenario, $\Nninety$ is the minimum number of stacked transits for which, if the modeled atmosphere is present, a featureless transmission spectrum would be rejected in at least 90\% of repeated observations. We use a Gaussian-equivalent $3\sigma$ false-alarm threshold for the primary calculation and retain $5\sigma$ as a conservative sensitivity test. The ratio $\Nninety/N_{\rm MCS}$ is our main diagnostic; a value of two means that the specified atmosphere requires twice the published MCS number of transits. This calculation provides an efficient physical screening step for identifying the planet-atmosphere combinations that merit full retrieval forecasts.

\subsection{Ariel small-planet sample}\label{subsec:sample}

The MCS repository provides separate files for known planets and TESS Planet Candidates (TPCs) \citep{MCSRepository}. We use only the known-planet release \texttt{Ariel\_MCS\_Known\_2026-05-11.csv}; no entries are drawn from the companion TPC catalog. We operationally define ``small planets'' by radius alone, $1.5 \leq R_{\rm p}/R_\oplus < 4.0$. This interval includes planets commonly described as super-Earths and sub-Neptunes; the label does not assume a bulk or atmospheric composition. We retain entries with finite, positive planetary radius, mass, temperature, stellar radius, and published MCS Tier-2 transit requirement. Stellar effective temperatures and $J$-band magnitudes, needed for the wavelength-dependent noise curves, are also required. After duplicate names are removed, these cuts yield 200 known planets. 

Because a planet's gravity directly controls its atmospheric scale height, and hence the amplitude of features in the transmission spectrum, we additionally quality-check masses that are clearly inconsistent with the current literature. We replace the catalog values for K2-138~f, Kepler-25~b, and Kepler-80~b with dynamical masses of 5.75, 10.8, and 6.93~$M_\oplus$, respectively \citep{Leleu2026K2138,Migaszewski2018,MacDonald2016}. Kepler-113~c, Kepler-176~c, and Kepler-33~c are excluded because no usable mass could be established for this calculation. The final sample therefore contains 197 planets. Other catalog masses are retained as published; their uncertainties are not propagated in the primary analysis.

Each target has an MCS estimate for the number of transits required for AIRS to detect a five-scale-height signal at ${\rm S/N}>7$, assuming an H$_2$/He-dominated atmosphere with mean molecular weight $\mu=2.3$ \citep{Edwards2019,MCSRepository}. For each of the final 197 planets, we calculate six hypothetical atmospheres, giving 1,182 planet--atmosphere cases. Here, we define a case as one planet evaluated under one specified atmosphere.

\subsection{Atmospheric forward models}\label{subsec:forward}

We generate transmission spectra with \prt{} v3.3.3 \citep{Molliere2019,Molliere2020}. The primary grid is the Cartesian product of three atmospheric metallicities and two cloud prescriptions. ``$1\times$ solar'' means that heavy elements have the solar abundance relative to hydrogen and helium; ``$10\times$'' and ``$100\times$ solar'' multiply that heavy-element abundance by 10 and 100 while the C/O ratio is held fixed at 0.55. These labels describe atmospheric enrichment and do not imply Earth-like composition. Chemical abundances and the pressure-dependent mean molecular weight are calculated in equilibrium for each case.

For every metallicity we calculate a ``clear'' spectrum and a ``1-mbar cloud'' spectrum. Clear means that no opaque gray cloud top is imposed. In the cloudy case, a fully opaque gray deck at $10^{-3}$ bar blocks transmission through deeper layers. This is an idealized sensitivity test, not a microphysical cloud prediction. The six primary cases are therefore $1\times$, $10\times$, and $100\times$ solar, each evaluated once without the opaque deck and once with the deck at 1 mbar.

The grid isolates two competing effects of enrichment. Increasing metallicity can strengthen molecular opacity, but it also increases the mean molecular weight and decreases the atmospheric scale height.
Detectability therefore does not vary monotonically with metallicity.

Each model spectrum is generated at higher resolution and binned to the 47 AIRS Tier-2 channels used by the MCS requirement: 36 AIRS-CH0 bins and 11 AIRS-CH1 bins. Restricting the calculation to AIRS ensures that the physical-spectrum test and MCS calibration use the same instrument. The complete opacity, chemistry, pressure-grid, cloud, and stellar-spectrum settings are provided in Appendix~\ref{app:forward_configuration}.

\subsection{Spectrally resolved noise}\label{subsec:noise}

For each target, we construct a wavelength-dependent one-transit uncertainty spectrum. The relative photon-noise shape follows a PHOENIX spectral energy distribution evaluated at the target star's stellar effective temperature and scaled with the $J$ magnitude, together with the AIRS channel widths. Following the simplified prescription used by \citet{Radica2026}, we combine the photon term in quadrature with a wavelength-independent contribution of 44.7\,ppm, obtained from a noise floor of 20\,ppm and a gain noise of 40\,ppm \citep{greene2016}. Although derived for JWST detectors, we adopt this value as a simple proxy for the non-photon contribution expected for a space-based infrared instrument such as Ariel; it is not a detector-by-detector systematics simulation.


We then adjust the photon-noise normalization separately for each target so that its clear $1\times$-solar reference spectrum reaches ${\rm S/N}=7$ after the published MCS number of Tier-2 transits. This anchors the calculation to the MCS sensitivity while retaining target-specific spectral weighting. Uncertainties are assumed to decrease as $N^{-1/2}$ when data from $N$ transits are combined. The method is therefore MCS-calibrated rather than an independent ArielRad calculation \citep{Mugnai2020}. Absolute transit counts should be read as internally consistent planning estimates, and the near agreement of the $1\times$-solar clear case with the MCS is essentially by construction.

Figure~\ref{fig:noise_curves} shows representative one-transit uncertainty spectra. Their absolute levels differ because of the target-by-target MCS calibration, while their wavelength dependence reflects the stellar spectra, AIRS channel widths, and the adopted fixed contribution. The discontinuity near $3.9\,\micron$ marks the boundary between AIRS-CH0 and AIRS-CH1.

\subsection{Tier-2 feasibility calculation}\label{subsec:feasibility}

For each physical spectrum we test the null hypothesis that the transmission spectrum is featureless, meaning that all 47 AIRS channels share one fitted constant transit depth. Assuming Gaussian noise, the statistic follows a $\chisq$ distribution with 46 degrees of freedom. Under the atmospheric model the statistic follows a noncentral-$\chisq$ distribution, and its noncentrality increases linearly with the number of stacked transits.

We define $\Nninety(3\sigma)$ as the smallest integer number of transits for which, if the modeled atmosphere is present, the featureless null would be rejected at the Gaussian-equivalent $3\sigma$ false-alarm threshold in at least 90\% of repeated observations. The rejection probability is evaluated analytically from the non-central-$\chisq$ distribution. We define $\Nninety(5\sigma)$ in the same way at the more conservative $5\sigma$ threshold. This is an atmosphere-detection calculation, not a molecular retrieval: rejecting a constant spectrum means that AIRS detects wavelength-dependent modulation, not that a particular molecule has been identified.

Our main comparison is the continuous requirement ratio
\begin{equation}
{\cal R}_{\rm MCS}=\frac{\Nninety(3\sigma)}{N_{\rm MCS}},
\label{eq:mcs_ratio}
\end{equation}
where $N_{\rm MCS}$ is the published Tier-2 requirement for the same planet. Values below one require fewer transits than the MCS reference, while values above one require more. We also quote the number of cases with $\Nninety\leq20$ as an illustrative summary, but the upper limit of 20 transits is not an Ariel selection rule or a physical boundary.

\section{Results}\label{sec:results}

\subsection{Physical requirements relative to the MCS reference}\label{subsec:population}

\begin{figure*}
\centering
\includegraphics[width=0.98\textwidth]{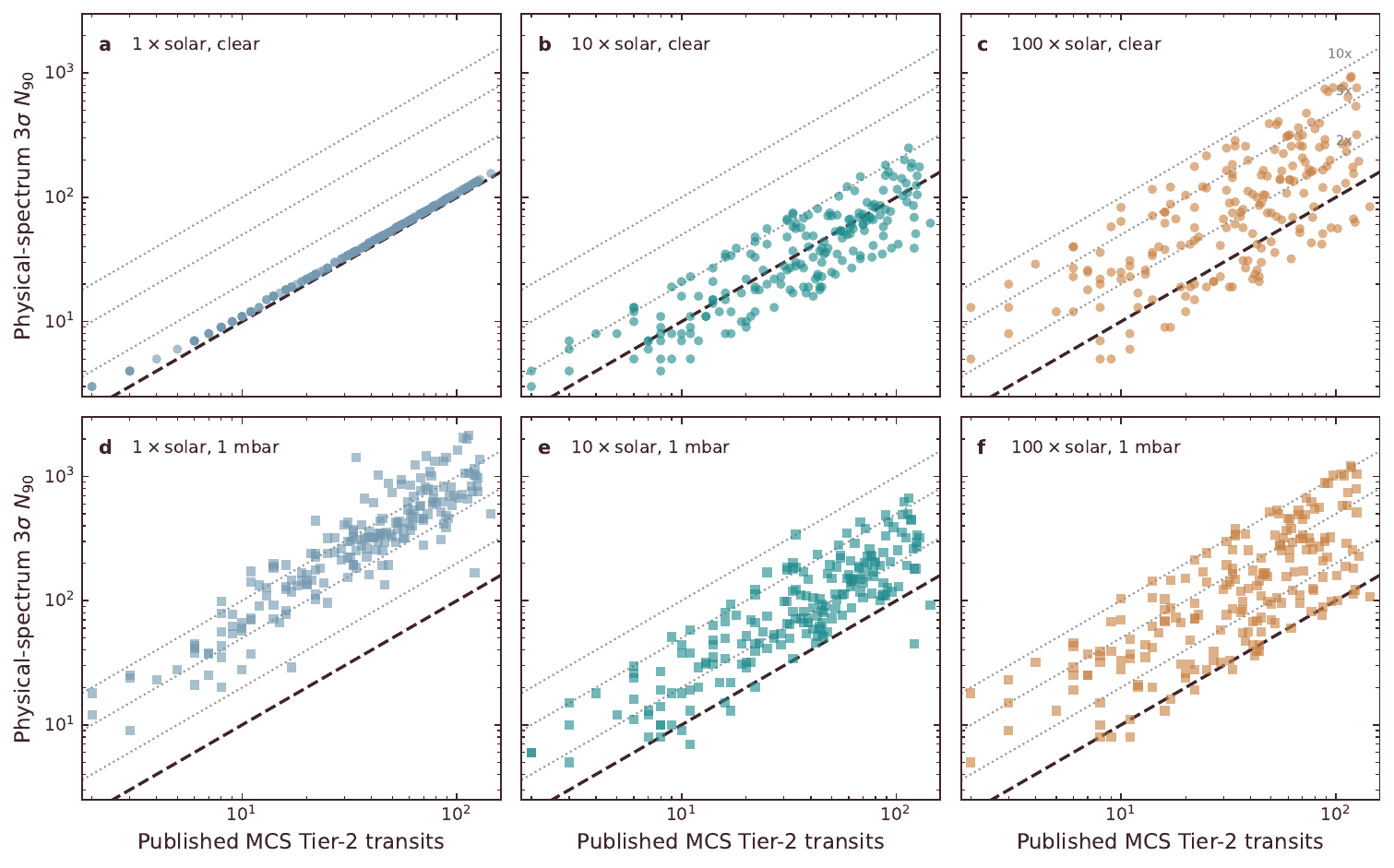}
\caption{Exact AIRS-only $3\sigma$ $\Nninety$ from the physical spectra compared with the published MCS Tier-2 requirement. Each panel shows one atmospheric scenario for all 197 planets. The dashed line marks equality; dotted lines mark factors of two, five, and ten above the MCS reference. The close agreement for the $1\times$-solar clear case is partly imposed by the MCS calibration.}
\label{fig:mcs_physical}
\end{figure*}

Figure~\ref{fig:mcs_physical} compares the exact physical-spectrum requirement with the published MCS Tier-2 requirement for every planet and atmospheric scenario. A planet on the one-to-one line needs the same number of transits as the MCS reference; planets above the line require more. The $1\times$-solar clear case lies close to the line by design because it is the spectrum used to normalize the target-specific noise. Its median requirement ratio is ${\cal R}_{\rm MCS}=1.09$, with a narrow 16th--84th percentile range of 1.08--1.13.

The other five scenarios show much broader and atmosphere-dependent departures. For clear atmospheres, the median ${\cal R}_{\rm MCS}$ values are 0.94 at $10\times$ solar and 2.40 at $100\times$ solar. For the 1-mbar cloud cases, the corresponding medians are 7.67, 2.33, and 3.51 at $1\times$, $10\times$, and $100\times$ solar. Table~\ref{tab:ratio_summary} gives the median and central 68\% interval for every case. Across the complete grid, 50.0\% of the cases require more than twice the MCS number of transits, 24.9\% require more than five times, and 4.7\% require more than ten times. Thus, in most atmospheric scenarios tested, small planets require more transits to robustly detect spectral features than listed in the MCS.

\begin{table}
\centering
\caption{Physical-to-MCS requirement ratio by atmospheric case. The interval contains the 16th--84th percentiles across the 197 planets.}
\label{tab:ratio_summary}
\small
\begin{tabular}{@{}lcc@{}}
\toprule
Case & Median ${\cal R}_{\rm MCS}$ & 16--84\% \\
\midrule
$1\times$ clear     & 1.09 & 1.08--1.13 \\
$10\times$ clear    & 0.94 & 0.53--1.80 \\
$100\times$ clear   & 2.40 & 0.79--5.29 \\
$1\times$, 1 mbar   & 7.67 & 5.49--12.05 \\
$10\times$, 1 mbar  & 2.33 & 1.43--4.30 \\
$100\times$, 1 mbar & 3.51 & 1.28--7.54 \\
\bottomrule
\end{tabular}
\end{table}

\subsection{An illustrative 20-transit summary}\label{subsec:changes}

Although the continuous ratio is the primary result, a common illustrative budget is useful for summarizing the grid. At $3\sigma$, 195 of the 1,182 planet--atmosphere cases require no more than 20 transits to robustly detect spectral features; at $5\sigma$, 106 do. Because each planet contributes six atmospheric cases, these counts summarize how the assumed atmosphere changes observational feasibility.

The $3\sigma$ counts for $1\times$, $10\times$, and $100\times$ solar are 52, 65, and 29 for clear atmospheres, and 4, 28, and 17 for the 1-mbar cloud cases (Table~\ref{tab:scenario_counts}). The peak at $10\times$ solar shows that detectability is not monotonic with metallicity: stronger molecular absorption can initially compensate for the decrease in scale height. 

\begin{table}
\centering
\caption{Planet--atmosphere cases for which no more than 20 transits are required to reject a featureless spectrum in at least 90\% of repeated observations at the listed false-alarm threshold.}
\label{tab:scenario_counts}
\small
\begin{tabular}{@{}lcc@{}}
\toprule
Case & $3\sigma$ & $5\sigma$ \\
\midrule
$1\times$ clear     & 52/197 & 33/197 \\
$10\times$ clear    & 65/197 & 38/197 \\
$100\times$ clear   & 29/197 & 12/197 \\
$1\times$, 1 mbar   & 4/197  & 2/197  \\
$10\times$, 1 mbar  & 28/197 & 14/197 \\
$100\times$, 1 mbar & 17/197 & 7/197  \\
\bottomrule
\end{tabular}
\end{table}

\subsection{Targets robust across the primary grid}\label{subsec:survivors}
TOI-2076~c and GJ~1214~b are the only planets for which all six primary atmospheres require no more than 20 transits to robustly detect spectral features at $3\sigma$. Their best, median, and worst primary-grid requirements are 3, 5, and 18 transits for TOI-2076~c, and 3, 9, and 18 transits for GJ~1214~b. Neither satisfies the same illustrative limit in all six cases at $5\sigma$: TOI-2076~c does so in five and GJ~1214~b in four.

Appendix~\ref{app:stress} extends these two planets to a 0.1-mbar cloud deck. Both exceed 20 transits in at least one of the additional cases, demonstrating that their apparent robustness depends on the maximum cloud altitude considered.

\section{Discussion}\label{sec:discussion}

\subsection{What the MCS reference captures}\label{subsec:fiveh}

The MCS five-scale-height calculation provides a transparent common scale for comparing heterogeneous targets. Our $1\times$-solar clear results closely track that reference, as expected from the noise normalization. The published MCS value therefore provides the baseline against which each enriched or cloudy atmospheric hypothesis can be evaluated.

Atmospheric scale height is nevertheless not the full observing statistic. AIRS measures a spectrum with wavelength-dependent noise and resolving power, while composition and clouds determine both the amplitude and placement of the absorption bands. The multiplier ${\cal R}_{\rm MCS}$ is consequently more useful than a universal correction: it states directly how much the requirement changes for the specified physical spectrum. The broad distributions for enriched and cloudy cases show that no single multiplicative adjustment applies to all small planets.

\subsection{How metallicity and clouds change the transit requirement}\label{subsec:metallicity}
The number of transits needed to robustly detect spectral features does not vary monotonically with metallicity. From $1\times$ to $10\times$ solar, the increase in molecular opacity can outweigh the decrease in scale height, producing larger spectral features in the Ariel passband. At $100\times$ solar, the mean-molecular-weight penalty often dominates. The exact balance depends on temperature, gravity, chemical equilibrium, and the placement of molecular bands relative to the noise curve.

Cloud-top pressure has a more uniformly adverse effect because a high-altitude opaque deck removes deeper atmospheric annuli that contribute to transmission modulation. The especially large median multiplier for the $1\times$-solar, 1-mbar case follows from combining a relatively weak equilibrium spectrum with an imposed high cloud. The 0.1-mbar calculation in Appendix~\ref{app:stress} quantifies the additional suppression produced by a still higher cloud deck.

GJ~1214~b provides an observational illustration of this sensitivity. Its HST/WFC3 transmission spectrum is nearly featureless and rules out several cloud-free compositions, requiring high-altitude aerosols \citep{Kreidberg2014}. JWST/MIRI phase-curve observations favor a high-metallicity atmosphere beneath a thick, reflective cloud or haze layer \citep{Kempton2023}, while JWST/NIRSpec transmission data show tentative CO$_2$ and CH$_4$ signatures above the aerosols \citep{Schlawin2024}. Joint HST--JWST modeling further favors an extremely metal-rich, possibly metal-dominated atmosphere \citep{Ohno2025}. Together, these observations disfavor clear, low-metallicity atmospheres as realistic descriptions of GJ~1214~b and place this well-studied sub-Neptune in the cloudy, highly enriched region represented by our models. For such atmospheric states, our calculations imply an Ariel requirement of approximately 2-10 times the published MCS estimate.

\subsection{Upper- and lower-atmosphere context from metastable helium}
\label{subsec:helium_context}

\begin{figure*}
\centering
\includegraphics[width=0.98\textwidth]
{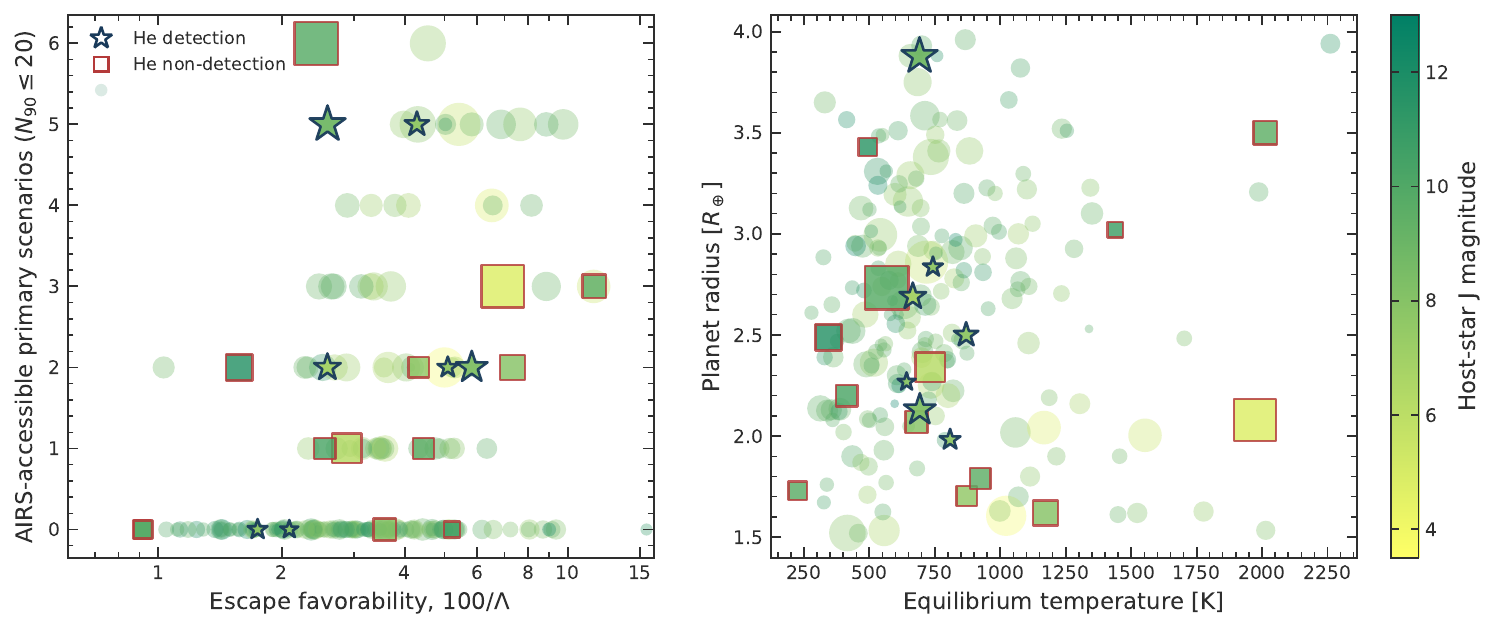}
\caption{Metastable-helium context for the 197 Ariel small planets. Left: $100/\Lambda$, a bulk escape-favorability proxy for which larger values indicate less strongly bound atmospheres, plotted against the number of the six primary atmospheric scenarios that satisfy the illustrative 20-transit guide. The six cases are $1\times$, $10\times$, and $100\times$ solar metallicity, each with either a clear atmosphere or an opaque 1-mbar cloud deck. The vertical axis ranges from 0 (no scenario can reject a featureless spectrum at $3\sigma$ in at least 90\% of repeated observations within 20 transits) to 6 (all scenarios can). Right: planetary radius versus equilibrium temperature. Stars with navy-blue outlines denote reported helium detections, squares with red outlines denote non-detections, and circles without outlines denote targets for which no published helium observation was identified. Marker color gives the host-star $J$ magnitude, and marker area encodes the standard \citet{Kempton2018} TSM. All planetary and host-star parameters are taken from the MCS tables.}
\label{fig:helium_context}
\end{figure*}

Featureless lower-atmosphere transmission spectra are not necessarily uninformative. For a small planet, weak AIRS spectral modulation may result from a high-mean-molecular-weight atmosphere, high-altitude clouds or hazes, or a combination of these effects. However, the molecular spectrum alone cannot establish whether the planet also possesses an extended, escaping upper atmosphere. Metastable helium \citep{OklopcicHirata2018} provides a complementary diagnostic that can help distinguish between these possibilities.

AIRS molecular transmission spectra and metastable-helium observations probe different vertical regions of an atmosphere. AIRS measures wavelength-dependent structure in the comparatively deeper molecular atmosphere, whereas the He\,I triplet at $1.0833\,\micron$ traces rarefied gas in the thermosphere and planetary wind and can reveal ongoing atmospheric escape \citep{OklopcicHirata2018,Spake2018,krishnamurthycowan2024}. Combining these measurements can therefore connect the planet's present-day molecular composition and cloud structure to the loss of its upper atmosphere. In particular, helium absorption can reveal an extended low mean molecular weight atmosphere even when the lower-atmosphere spectrum is muted, although a helium non-detection does not by itself demonstrate the absence of atmospheric escape \citep{Zhang2025}. Reported helium detections therefore demonstrate that some small planets retain extended H$_2$/He atmospheres even while those atmospheres are being lost.

We crossmatched the final 197-planet sample against a literature census of published metastable-helium observations, frozen on 2026 August 25. The resulting sample contains 7 reported helium detections, 13 non-detections, and 177 targets for which no published helium observation was identified.

As a simple bulk measure of how strongly an atmosphere is gravitationally bound, we calculate the restricted Jeans parameter \citep{Fossati2017},
\begin{equation}
\Lambda =
\frac{G M_{\rm p}m_{\rm H}}
     {k_{\rm B}T_{\rm eq}R_{\rm p}},
\label{eq:restricted_jeans}
\end{equation}
where $G$ is the gravitational constant, $M_{\rm p}$ and $R_{\rm p}$ are the planetary mass and radius, $m_{\rm H}$ is the mass of a hydrogen atom, $k_{\rm B}$ is the Boltzmann constant, and $T_{\rm eq}$ is the planetary equilibrium temperature reported in the MCS. We use the quality-controlled planetary masses adopted in the main analysis. The dimensionless parameter $\Lambda$ compares the gravitational binding energy of a hydrogen atom with its thermal energy; smaller values indicate less strongly bound atmospheres. The left panel of Fig.~\ref{fig:helium_context} displays $100/\Lambda$, where the factor of 100 is included only for visual convenience. Larger values indicate atmospheres that are less strongly bound according to this bulk metric. The vertical axis is an integer count from zero to six and gives the number of the six primary atmospheric scenarios for which no more than 20 transits are required to reject a featureless spectrum at the $3\sigma$ threshold in at least 90\% of repeated observations. Thus, zero means that none of the six scenarios satisfies the illustrative 20-transit guide, whereas six means that all six do.

The right panel places the same targets in planetary-radius--equilibrium-temperature space. Marker area encodes the standard $J$-band Transmission Spectroscopy Metric (TSM) of \citet{Kempton2018},
\begin{equation}
{\rm TSM} =
S\,\frac{R_{\rm p}^{3}T_{\rm eq}}
          {M_{\rm p}R_{\star}^{2}}\,
10^{-J/5},
\label{eq:kempton_tsm}
\end{equation}
where $R_{\rm p}$ and $M_{\rm p}$ are in Earth units, $R_\star$ is in solar units, and $J$ is the host-star $J$-band magnitude. For the radius range considered here, the scale factor is $S=1.26$ for $1.5\leq R_{\rm p}/R_\oplus<2.75$ and $S=1.28$ for $2.75\leq R_{\rm p}/R_\oplus<4.0$ \citep{Kempton2018}. The host-star $J$ magnitude is also shown by marker color.

The helium detections and non-detections overlap substantially in both panels, and neither $100/\Lambda$ nor the standard TSM cleanly separates them. This overlap is expected because metastable-helium absorption depends on the stellar XUV spectrum, photoionization and recombination, atmospheric composition, wind structure, and temporal variability in addition to the bulk planetary properties \citep{OklopcicHirata2018,krishnamurthycowan2024,krishnamurthy_w107bhelium}. The TSM was designed as a broadband molecular-transmission observability metric rather than a predictor of narrow-band helium detectability. 

The crossmatch adds a second science-prioritization dimension. Ariel targets with reported helium absorption demonstrate that some small planets retain extended H$_2$/He atmospheres while actively losing them. Tier-2 molecular spectra of these planets would connect atmospheric escape to the composition and cloud structure below. Conversely, AIRS-favorable planets without published helium measurements provide natural targets for complementary upper-atmosphere observations. A joint upper--lower-atmosphere sample would therefore add direct evolutionary context to Ariel's small-planet survey.

\subsection{Implications for Ariel target selection}\label{subsec:implications}

An Ariel Tier-2 survey of small planets should begin with a population-level scientific question:for example, how atmospheric mean molecular weight and cloud properties vary across the sub-Neptune population; and then select a sample with sufficient spectral precision and parameter-space leverage to answer it. This is especially important for small planets because their relatively weak transmission signals often require many repeated observations. Recent Ariel sample-optimization studies have therefore restricted their candidate samples to planets requiring no more than 20 transits to reach Tier-2 precision \citep{Panek2026}. Our results show that atmospheric assumptions can materially alter which planets satisfy such an observing-time constraint: in every atmospheric scenario explored, most planets require more than 20 transits to robustly detect spectral features at $3\sigma$.

We therefore recommend a staged approach to small-planet target selection. The MCS can first provide a common instrumental reference from which to construct a science-driven shortlist. The shortlisted planets can then be evaluated using composition- and cloud-specific physical spectra, particularly when the scientific objective depends on distinguishing low- and high-mean-molecular-weight atmospheres or measuring the effects of clouds. The most important target--atmosphere combinations can subsequently be examined with retrieval-level forecasts to determine the expected constraints on molecular abundances and cloud properties. This refinement should follow, rather than replace, choices based on population leverage, observability, ephemeris quality, complementary upper-atmosphere information, and scientific priority.

The physical forward-model calculation developed here provides an efficient intermediate screening step. The planet-specific $\Nninety$ values and requirement multipliers quantify how the observing cost changes when plausible atmospheric states replace the common five-scale-height reference. They can therefore identify cases for which the nominal MCS allocation is likely to provide an informative spectrum, cases that require additional visits, and high-value targets for which detailed retrieval forecasts are warranted. The helium crossmatch supplies a complementary evolutionary criterion: planets with detected extended H$_2$/He atmospheres are particularly valuable for connecting atmospheric escape to the molecular composition and cloud structure that Ariel will probe below.

These results expose three possible strategies for an Ariel Tier-2 small-planet survey. The mission could exclude planets whose atmosphere-specific requirements are prohibitively expensive; retain a broad sample while accepting that some spectra may remain weakly constraining; or concentrate substantially more observations, including more than 20 transits where scientifically justified, on a smaller sample to obtain robust atmospheric constraints. The appropriate balance depends on the population-level scientific question. The atmosphere-specific requirements reported here quantify the observational cost of each strategy and provide a practical basis for designing an informative Ariel survey of small planets.

\subsection{Scope and limitations}\label{subsec:limitations}

This work measures the detectability of wavelength-dependent spectral modulation. Full retrieval forecasts can extend the analysis to molecular-abundance precision and degeneracies for the highest-priority targets. The present grid assumes equilibrium chemistry and idealized gray clouds and omits disequilibrium chemistry, wavelength-dependent aerosol opacity, stellar contamination, time-correlated instrumental noise, and uncertainties in planetary mass and radius.

Because the spectral noise shape is calibrated to the published MCS requirements rather than generated through direct ArielRad simulations for every planet, the absolute $\Nninety$ values retain the MCS sensitivity normalization. A future batch ArielRad calculation would provide an external check of the absolute requirements and channel-to-channel covariance. Propagating the planetary mass and radius uncertainties will also be important because gravity directly controls atmospheric scale height.

\section{Conclusions}\label{sec:conclusions}

We have evaluated physical equilibrium-chemistry spectra for 197 known small planets in the Ariel MCS under six atmospheric scenarios and expressed the resulting AIRS requirements relative to the published MCS reference. Our principal conclusions are:
\begin{enumerate}
\item The MCS five-scale-height prescription is an effective common screening reference. The $1\times$-solar clear physical calculation has a median $\Nninety/N_{\rm MCS}=1.09$, with the close agreement partly imposed by our MCS calibration.
\item Atmospheric assumptions can change the required time substantially. Median requirement multipliers are 0.94 and 2.40 for $10\times$- and $100\times$-solar clear atmospheres, and 7.67, 2.33, and 3.51 for the $1\times$, $10\times$, and $100\times$-solar 1-mbar cloud cases.
\item Detectability is non-monotonic with metallicity because stronger molecular opacity competes with the smaller scale height of a heavier atmosphere.
\item Using 20 transits as an illustrative upper limit, 195 of 1,182 physical cases satisfy the primary $3\sigma$ criterion and 106 satisfy the $5\sigma$ test. In every atmospheric scenario explored, most planets require more than 20 transits to robustly detect spectral features.
\item GJ~1214~b and TOI-2076~c remain within 20 transits in all six primary cases at $3\sigma$, but neither remains within that limit in the presence of a 0.1-mbar cloud deck.
\item The helium crossmatch identifies systems that already show extended H$_2$/He upper atmospheres, providing high-value targets for connecting atmospheric escape with Tier-2 measurements of molecular composition and clouds.
\end{enumerate}
An Ariel Tier-2 small-planet survey must therefore balance sample size against atmospheric information: it can exclude the most expensive planets, retain a broader sample with some weakly constraining spectra, or observe a smaller sample deeply enough to obtain robust constraints. The atmosphere-specific multipliers reported here quantify the transit cost of these choices and identify where detailed retrieval forecasts will provide the greatest value.

\section*{Acknowledgments}

We thank the Ariel Mission Consortium and the developers of the Ariel Mission Candidate Sample and \texttt{petitRADTRANS}. This work used the MCS release dated 2026 May 11.
Software used in this work includes \prt{} v3.3.3 \citep{Molliere2019,Molliere2020}, Python, NumPy, SciPy, pandas, and Matplotlib.

\section*{Data Availability}

\section*{Conflict of Interest}
Authors declare no conflict of interest.

\bibliographystyle{rasti}
\bibliography{references}
\appendix

\section{Forward-model configuration}
\label{app:forward_configuration}

Table~\ref{tab:forward_model_settings} records the configuration used to generate the atmospheric transmission spectra and construct the relative wavelength dependence of the target-specific noise curves.

\begin{table*}
\centering
\caption{Atmospheric forward-model and stellar-spectrum configuration
adopted in this study.}
\label{tab:forward_model_settings}
\small
\renewcommand{\arraystretch}{1.15}
\begin{tabular}{@{}p{0.23\textwidth}p{0.71\textwidth}@{}}
\toprule
Setting & Adopted configuration \\
\midrule

Radiative-transfer code &
\texttt{petitRADTRANS} v3.3.3 in transmission geometry
\citep{Molliere2019,Molliere2020}. \\

Native wavelength range &
$0.5$--$7.8\,\micron$. Line absorption was calculated with
correlated-$k$ opacity tables at $R=1000$. \\

Pressure grid &
100 layers uniformly spaced in $\log P$ from $10^{-6}$ to
$10^{2}$ bar. \\

Temperature structure &
Isothermal at the planetary temperature reported in the MCS for each
target. \\

Chemistry &
Chemical-equilibrium abundances interpolated with the
\texttt{PreCalculatedEquilibriumChemistryTable} supplied with
\texttt{petitRADTRANS}. The C/O ratio was fixed to 0.55, and the
atmospheric metallicity was set to $1\times$, $10\times$, or
$100\times$ solar. The pressure-dependent mean molecular weight was
taken from the equilibrium-chemistry calculation. \\

Line-opacity species &
H$_2$O (POKAZATEL), CO with natural isotopic abundances (HITEMP),
CO$_2$ (UCL-4000), CH$_4$ (HITEMP), and NH$_3$ (CoYuTe). \\

Continuum opacity &
H$_2$--H$_2$ and H$_2$--He collision-induced absorption. \\

Rayleigh scattering &
H$_2$ and He. \\

Reference radius and gravity &
The cataloged planetary radius was assigned to a reference pressure
of $10^{-2}$ bar. The reference gravity was calculated as
$g=GM_{\rm p}/R_{\rm p}^{2}$ using the MCS radius and the quality-controlled mass described in Section~\ref{subsec:sample}. \\

Cloud treatment &
Either no opaque cloud deck (``clear'') or an idealized gray,
fully opaque cloud top at $10^{-3}$ bar (1 mbar). The additional
stress test used a cloud top at $10^{-4}$ bar (0.1 mbar) for
GJ~1214~b and TOI-2076~c only. \\

Stellar spectra &
Stellar spectral shapes were obtained from the PHOENIX table supplied
with \texttt{petitRADTRANS} and evaluated at the MCS stellar effective
temperature. Each spectrum was normalized at the $J$-band pivot
wavelength of $1.235\,\micron$ and scaled using the cataloged
$J$-band magnitude. The stellar spectra determine the relative
wavelength dependence of the photon-noise model. \\

AIRS binning &
The spectra were binned into 47 wavelength channels: 36 AIRS-CH0 bins from
$1.95$--$3.90\,\micron$ and 11 AIRS-CH1 bins from
$3.90$--$7.80\,\micron$. The exact test therefore has 46 degrees of freedom after fitting one constant transit depth. \\

\bottomrule
\end{tabular}
\end{table*}


\section{High-altitude cloud stress test}\label{app:stress}

\begin{figure*}
\centering
\includegraphics[width=0.94\textwidth]{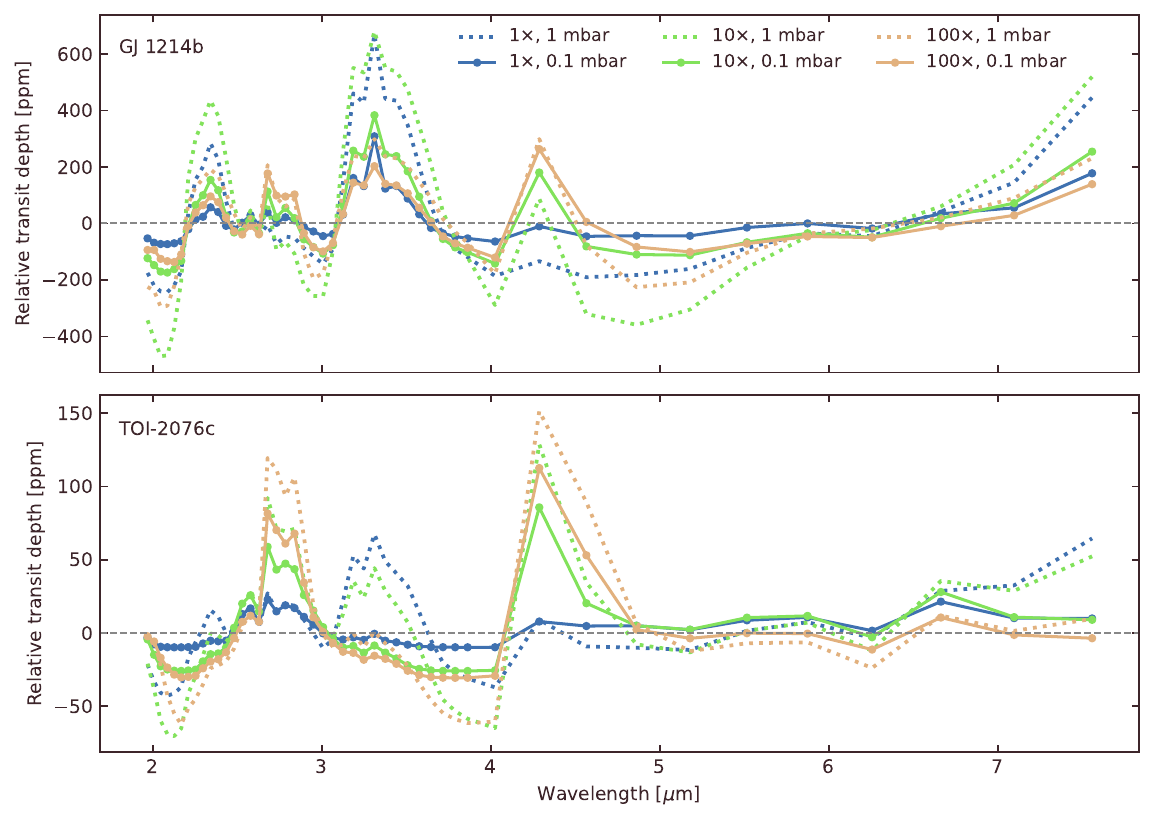}
\caption{AIRS-binned transmission spectra for the two targets that satisfy the 20-transit guide in all six primary cases. Dotted curves show the primary 1-mbar cloud cases and solid curves show the 0.1-mbar stress test. The higher cloud generally suppresses wavelength-dependent modulation.}
\label{fig:cloud_spectra}
\end{figure*}

The six primary scenarios identify GJ~1214~b and TOI-2076~c as the only planets with $\Nninety\leq20$ in every case at $3\sigma$ (Section~\ref{subsec:survivors}). To test how that statement depends on the cloud range, we recompute these two planets at $1\times$, $10\times$, and $100\times$ solar metallicity with an opaque gray cloud top at $10^{-4}$ bar (0.1 mbar). All other forward-model settings, AIRS binning, target-specific noise curves, and the $\Nninety$ calculation are unchanged. This targeted extension is sufficient for the stated conditional question because the other 195 planets had already exceeded 20 transits in at least one primary scenario.

For GJ~1214~b, the \citet{Kreidberg2014} posterior permits lower cloud-top pressures as the mean molecular weight increases. JWST observations favor a highly enriched atmosphere with $\mu\gtrsim10$, making cloud tops near $10^{-6}$--$10^{-5}$ bar plausible \citep{Kempton2023,Schlawin2024,Ohno2025}. The adopted 0.1-mbar deck is therefore an intermediate rather than an extreme stress test.

Moving the cloud top from 1 mbar to 0.1 mbar removes a larger fraction of the atmospheric annulus and generally suppresses the wavelength-dependent transmission signal (Fig.~\ref{fig:cloud_spectra}). The response remains non-monotonic in metallicity because molecular opacity, mean molecular weight, and AIRS noise weighting compete.
\begin{figure*}
\centering
\includegraphics[width=0.94\textwidth]{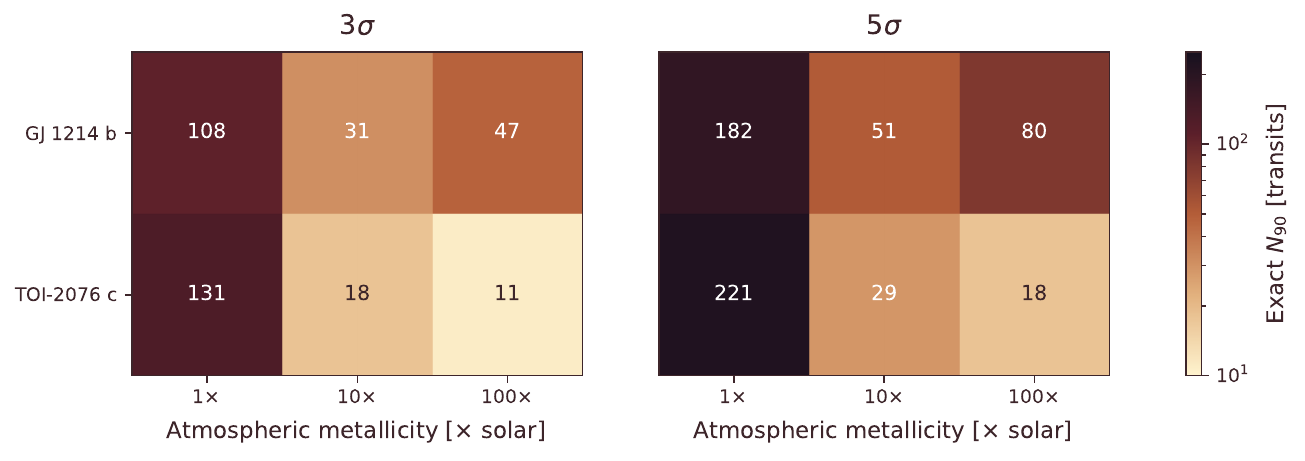}
\caption{AIRS-only 0.1-mbar cloud stress test for the two targets that satisfy the 20-transit guide in all six primary cases. Each value is $\Nninety(3\sigma)$: the minimum number of transits required to reject a featureless spectrum at the $3\sigma$ false-alarm threshold in at least 90\% of repeated observations. The most favorable metallicity differs between the targets (i.e., $10\times$ and $100\times$ solar for GJ~1214~b and TOI-2076~c respectively), emphasizing the non-monotonic interaction of opacity, scale height, and wavelength-dependent noise weighting.}
\label{fig:stress_heatmap}
\end{figure*}

The exact transit requirements are shown in Fig.~\ref{fig:stress_heatmap}. GJ~1214~b satisfies the 20-transit guide in six of the nine extended scenarios and has a worst-case $3\sigma$ $\Nninety$ of 108. TOI-2076~c satisfies it in eight of nine and has a worst-case $\Nninety$ of 131; its $10\times$- and $100\times$-solar 0.1-mbar cases require 18 and 11 transits, respectively. Thus, neither planet remains below 20 transits throughout the complete nine-scenario grid. The numerical values are printed in the heatmap, so no separate table is needed.

\bsp
\label{lastpage}
\end{document}